\documentclass[11pt]{article}

\usepackage{amsmath,amsfonts,amssymb,amsthm,bbm,stmaryrd}
\usepackage[dvips]{graphicx}
\usepackage{latexsym}
\usepackage{psfrag}
\usepackage[latin1]{inputenc}
\usepackage{hyperref}
\usepackage{color}
\usepackage{enumerate}
\numberwithin{equation}{section}

\newcommand{\be}{\begin{equation}}
\newcommand{\ee}{\end{equation}}
\newcommand{\barray}{\begin{array}}
\newcommand{\earray}{\end{array}}
\newcommand{\bea}{\begin{eqnarray}}
\newcommand{\eea}{\end{eqnarray}}
\newcommand{\bs}{\begin{subequations}}
\newcommand{\es}{\end{subequations}}

\newcommand{\bit}{\begin{itemize}}
\newcommand{\eit}{\end{itemize}}
\newcommand{\bd}{\begin{description}}
\newcommand{\ed}{\end{description}}

\def\nn{\nonumber}

\def\la{\langle}
\def\ra{\rangle}

\newcommand{\mean}[1]{\la{#1}\ra}

\newcommand{\vet} [2] {\left ( \begin{array}{c}{#1}\\{#2} \end{array} \right ) }
\newcommand{\mat} [4] {\left ( \begin{array}{cc}{#1}&{#2}\\{#3}&{#4} \end{array} \right ) }

\def\w{\wedge}
\newcommand{\p}{\partial}
\newcommand{\na}{\nabla}

\newcommand{\f}{\frac}
\newcommand{\tl}{\tilde}

\renewcommand{\a}{\alpha} \renewcommand{\b}{\beta} \newcommand{\g}{\gamma}  
\renewcommand{\d}{\delta}  \newcommand{\eps}{\epsilon} 
 \renewcommand{\th}{\theta}  \newcommand{\vth}{\vartheta} 
    \renewcommand{\l}{\lambda}
\let\m=\mu    \let\n=\nu   \let\r=\rho \let\om=\omega
 \newcommand{\s}{\sigma}     \let\vphi=\varphi 
\let\G=\Gamma \let\D=\Delta  \let\Th=\Theta  
\let\Si=\Sigma \let\Om=\Omega

\def\cH{{\cal H}}

\newcommand{\eqSi}{\stackrel{\Si}=}
\newcommand{\pb}[1]{\stackrel{{#1}}=}

\newcommand{\sscr}{\scriptscriptstyle\rm}

\newcommand{\qB}{{\bar q}}
\newcommand{\nB}{{\bar n}}
\newcommand{\uB}{{\bar u}}
\newcommand{\KB}{{\bar K}}
\newcommand{\ssL}{{\sscr L}}

\newcommand{\Nc}{N_{\sscr S}}

\DeclareMathOperator\arctanh{arctanh}

\begin{document}

\title{\bf Brown-York charges with mixed boundary conditions}

\author{\Large{Gloria Odak and Simone Speziale}
\smallskip \\ 
\small{\it{Aix Marseille Univ., Univ. de Toulon, CNRS, CPT, UMR 7332, 13288 Marseille, France}} }
\date{\today}

\maketitle

\begin{abstract}
We compute the Hamiltonian surface charges of gravity for a family of conservative boundary conditions, that include Dirichlet, Neumann, and York's mixed boundary conditions defined by holding fixed the conformal induced metric and the trace of the extrinsic curvature. We show that for all boundary conditions considered, canonical methods give the same answer as covariant phase space methods improved by a boundary Lagrangian, a prescription recently developed in the literature and thus supported by our results.
The procedure also suggests a new integrable charge for the Einstein-Hilbert Lagrangian, different from the Komar charge for non-Killing and non-tangential diffeomorphisms. We study how the energy depends on the choice of boundary conditions, showing that both the quasi-local and the asymptotic expressions are affected.
Finally, we generalize the analysis to non-orthogonal corners, confirm the matching between covariant and canonical results without any change in the prescription, and discuss the subtleties associated with this case.
\end{abstract}

\tableofcontents

%--------------------------------------------------------------
\section{Introduction}
%--------------------------------------------------------------

\vspace{.5cm}

\begin{flushright}
\small{\emph{I keep warning you. Doors and corners, kid. That's where they get you.} \\ Miller}
\end{flushright}

It is a fundamental property of general relativity that energy is not conserved, but dissipated by gravitational radiation. 
A notion of conserved energy in phase space can be introduced if one restricts attention to non-radiative spacetimes. 
An example of conserved energy is the ADM charge \cite{Regge:1974zd} at spatial infinity, or its quasi-local version the Brown-York (BY) charge \cite{Brown:1992br} (see \cite{Szabados:2009eka} for a review on quasi-local charges). This notion of energy is however not universal, and depends on the way the system is made conservative, namely on the specific choice of boundary conditions used to eliminate dissipation. The ADM and BY formulas for instance, are based on Dirichlet boundary conditions.
In this paper we study how the value of the energy changes as we move from Dirichlet to York's mixed boundary conditions, to Neumann's. This corresponds to fixing less components of the induced metric, and more components of its momentum, namely the extrinsic curvature. We will see that the neat effect of this process 
is to produce smaller values of the energy. For instance applying the formula to the Kerr spacetime, we find respectively $M$, $2M/3$ and $M/2$.

To obtain these results, we use two different methods. The first is the covariant phase space.
It is particularly powerful because it allows one to treat also the radiative/dissipative case, and features prominently in the study of gravitational radiation (see e.g. \cite{Ashtekar:1981bq,Wald:1999wa,Barnich:2001jy,Barnich:2011mi,Compere:2020lrt,Freidel:2021yqe}). In the non-radiative/conservative case, the method can be used to obtain Hamiltonians studying how imposing vanishing symplectic flux at the boundary selects a specific symplectic potential leading to integrable charges (see e.g. discussion in \cite{Chandrasekaran:2020wwn,Freidel:2021cbc}). The idea of this construction was explained in \cite{Iyer:1994ys} (see also \cite{Barnich:2007bf} for a related interpretation) and made more explicit in \cite{Iyer:1995kg,Harlow:2019yfa}, where it was shown that for Dirichlet boundary conditions one obtains precisely  the BY charge. It was recently more systematically developed by Freidel, Geiller and Pranzetti (FGP) \cite{Freidel:2020xyx}, arriving at a general prescription for the charges that is valid for arbitrary boundary conditions.\footnote{The analysis was further extended in \cite{Chandrasekaran:2020wwn,Freidel:2021cbc} to include anomalies, but these will not play a role here.} The charge is defined in an unambiguous way, and depends on both the bulk Lagrangian and the boundary term required by the variational principle associated with the chosen boundary conditions, as anticipated in \cite{Compere:2008us}.
 The FGP prescription, given by equation \eqref{HamC} below, is thus a perfect tool to investigate the question raised above.

The second method is the straightforward canonical analysis based on the ADM decomposition, which has no problem in dealing with the non-radiative context. The calculation we present is a simple extension of the analysis done in \cite{Hawking:1996ww,Brown:2000dz} for Dirichlet boundary conditions. A nice feature of this extension is to see explicitly how the boundary term changes the kinetic part of the ADM Lagrangian to recast it in the form consistent with the symplectic potential associated with the chosen boundary conditions. 
We find that the canonical method reproduces exactly the same charges obtained with the FGP prescription in the covariant method, for all cases considered.
A consequence of our results is thus to offer support to the prescription of \cite{Freidel:2020xyx}.

These results are based on the simplest set-up with an orthogonal corner between the time-like boundary and the space-like hypersurfaces,
a situation where 3d boundary Lagrangians are sufficient to make the variational principle well-defined.
We also investigate a more general context with non-orthogonal corners.
 In the presence of non-orthogonal corners, the variational principle requires an additional 2d term \cite{Hartle:1981cf,Hayward:1993my} (see also \cite{Lehner:2016vdi,Jubb:2016qzt,Oliveri:2019gvm} for recent work). 
Since the rationale for constructing the covariant surface charges is to use an action with a well-posed variational principle, one may 
wonder if the corner Lagrangian contributes to the formula for the charges as well.

To address this question, we repeat the calculations using the FGP prescription with different boundary conditions, this time allowing for non-orthogonal corners, and compare the results with the ones obtained with canonical methods. Using canonical methods \cite{Hawking:1996ww,Brown:2000dz}, it is known that the presence of a non-orthogonal corner has both a physical and a mathematical consequence. The physical consequence is that one can consider two different classes of observers, those at rest with respect to the space-like foliation of spacetime, and those at rest along the time-like boundary.
At the corner, these are related by a boost transformation, making the different notions of energy directly comparable. The mathematical consequence is that the charges pick up `tilting terms', namely they depend explicitly on the boost between the normals. 
This dependence allows us a further, non-trivial test passed by the FGP prescription: The covariant and canonical results match also for non-orthogonal corners, without any amendment to the formula.
This matching is however subtle, because in order to obtain it, one needs to take into account that the boundary term contains a kinetic term that has to be independently put in Hamiltonian form via a Legendre transform \cite{Brown:2000dz,Harlow:2019yfa}. The role of the corner Lagrangian is only to adjust the boundary kinetic term to match the chosen boundary conditions.

\begin{figure}[h!]
\centering 
\includegraphics[width=7cm]{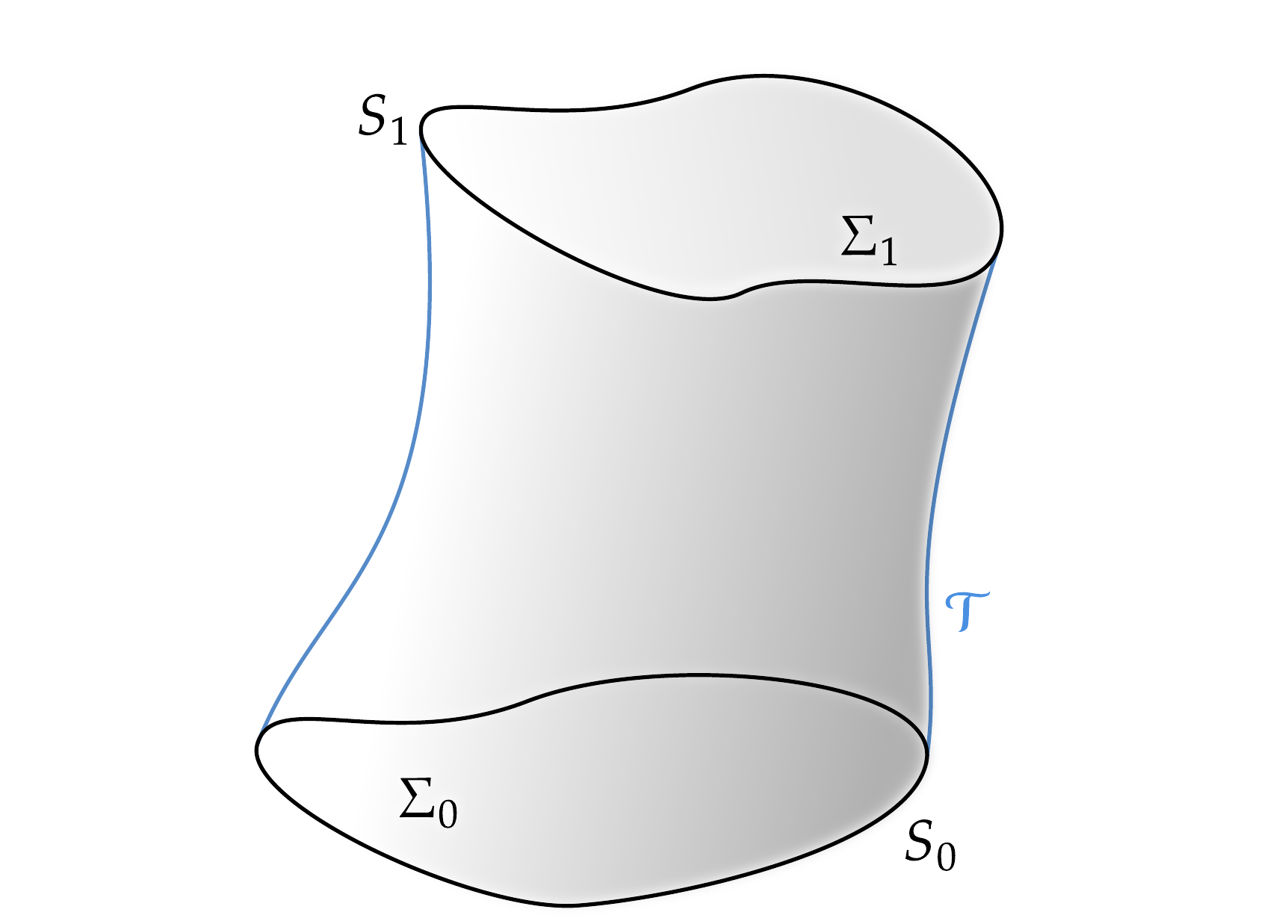} \hspace{2cm}
\includegraphics[width=5cm]{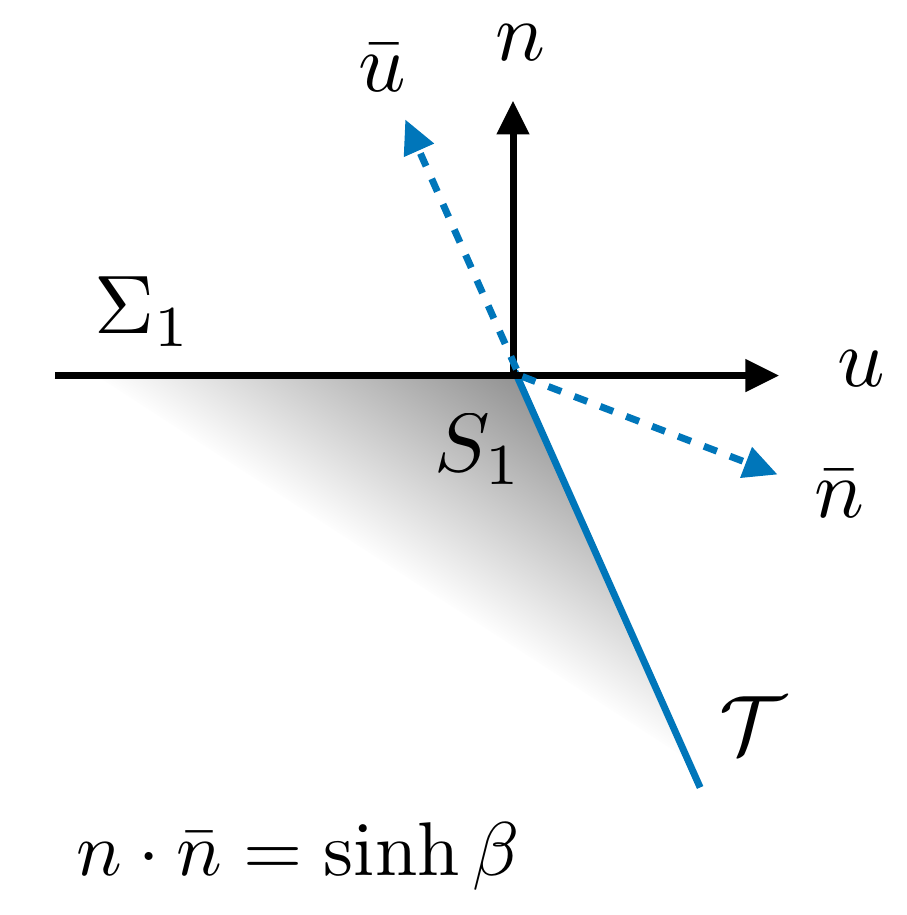}
\caption{\label{Fig1}
\small{
Left panel: \emph{A finite region of spacetime and the notation used for the boundaries. }
Right panel: \emph{The four normals generically present at the corner, here the one in the future of $\cal T$. Unbarred quantities refer to the space-like hypersurface, and barred quantities to the time-like one.} 
} }
\end{figure}

Throughout the paper we use mostly-plus metric signature, and greek letter for spacetime indices. The notation for the different boundaries and their geometric objects are summarized in Fig.~\ref{Fig1} and Table~\ref{tab:defn}.

 \begin{table}[htb]
 \begin{center}
  \begin{tabular}{l|c|ccccc}  
  & & metric & $s$ & normal & extrinsic curvature & boundary normal \\ \hline
   Spacetime & $M$ & $g$ & \\
   Space-like hypersurfaces & $\Sigma$ & $q$ & -1 & $n$ & $K$ & $u$ \\
   Time-like hypersurfaces & $\cal T$  & $\qB$ & 1 & $\nB$ & $\KB$  & $\uB$\\
%   Null hypersurfaces & $\cal N$ & $\qB$ & 0 & $\nB$ & $(\s,\th)$\\
  Corners & $S$ & $\g$ &  & $(n,u)$ or $(\nB,\uB)$ &  $k$ & \\ \hline  
  \end{tabular}
  \caption{\small{\emph{Notations for the different boundaries.}}  \label{tab:defn}}
 \end{center}   
 \end{table}

%--------------------------------------------------------------
\section{Mixed boundary conditions} \label{sec:mixed}
%--------------------------------------------------------------

We briefly review the different boundary conditions we will consider in this paper. 
We start from the Einstein-Hilbert (EH) Lagrangian,
\be
L^{\sscr EH} = R \eps,
\ee
where $\eps:=\sqrt{-g}d^4x$ is the volume 4-form (See Appendix~\ref{AppConv} for our conventions on volume forms and orientations), and we fix units $16\pi G=1$.
The variation gives 
$\d L^{\sscr EH} = G_{\m\n}\d g^{\m\n}\eps+d\th^{\sscr EH},$ where
\begin{align}
\th^{\sscr EH} =\f1{3!} \th^\m \eps_{\m\n\r\s}~dx^\n\w dx^\r \w dx^\s , \qquad
\th^\m = 2 g^{\r[\s} \d \G^{\m]}_{\r\s} = 2g^{\m[\r}g^{\n]\s} \na_\n\d g_{\r\s}. \label{thg}
\end{align}

Consider a hypersurface $\Si$, with $n_\m$ its unit normal, and boundary $\p\Si=S$, with $u_\m$ its unit normal within $T^*\Si$, so that $u\cdot n=0$. The corresponding volume forms are $\eps_\Si$ and $\eps_S$, related among each other and  to $\eps$ as in Appendix~\ref{AppA}. We treat in this Section both cases of space-like or time-like $\Sigma$ at once, and accordingly we define $s:=n^2=\mp 1=-u^2$. From the next Section, we will use a different notation for the different boundaries for the sake of clarity. 
We denote  $q_{\m\n}:=g_{\m\n} -s n_\m n_\n$ the projector, whose pull-back on $\Si$ gives the induced metric, with determinant $q$;  and $K_{\m\n}:=q^\r_\m\na_\r n_\n$ the extrinsic curvature, with $K:=g^{\m\n}K_{\m\n}=\na_\m n^\m$.
Taking the pull-back of the boundary variation on $\Si$ one has (see e.g. \cite{Brown:2000dz,Lehner:2016vdi,Oliveri:2019gvm})
\begin{align}\label{thEH}
\th^{\sscr EH} &\pb{\Si} s \left( K_{\m\n} \d q^{\m\n}-2\d K \right) \eps_\Si + d\vth^{\sscr EH} 
= s \, q_{\m\n}\d\tl\Pi^{\m\n} d^3x +d\vth^{\sscr EH}, \\
\vth^{\sscr EH}&:= - u_\m  \d n^\m \eps_S = u^\m n^\n \d g_{\m\n}  \eps_S.
\end{align}
In the second equality of \eqref{thEH} we introduced the gravitational momentum
\be
\tl\Pi^{\m\n}:=\sqrt{q}(K^{\m\n}-q^{\m\n}K),\qquad \tl\Pi:=g_{\m\n}\tl\Pi^{\m\n} = -2\sqrt q K,
\ee
familiar from the ADM analysis, here written as a spacetime tensor. We will favour covariant expressions throughout the paper, and avoid using  hypersurface indices.\footnote{On the other hand, one should keep in mind that the actual induced metric is the pull-back of the projector $q_{\m\n}$, so that variations which keep the induced metric fixed satisfy $q^\m_\r q^\n_\s \d q_{\m\n}\pb{\p M} \d q_{\m\n}\pb{\p M}0$, but not necessarily $\d q_{\m\n}=0$.}

The geometric decomposition \eqref{thEH} of the boundary variation allows us to study the variational principle in a finite region of spacetime like the one in Fig.~\ref{Fig1}. 
The variational principle is well-defined only if the total boundary contribution vanish, so that the action is correctly extremized on-shell.\footnote{The time-like and space-like parts of the boundary variation have different conceptual status. To define the phase space, it is enough to ensure that the time-like variation vanishes. Vanishing of the space-like variation will then correspond to picking a specific solution in the phase space.} 
Let us see when this happens.
Gluing \eqref{thEH} along the boundary, one obtains the corner variation \cite{Hayward:1993my}\footnote{See also \cite{Lehner:2016vdi} for a more recent derivation, and \cite{Jubb:2016qzt,Oliveri:2019gvm} to see how the derivation is simplified if one uses tetrad instead of metric variables.} 
\be\label{dbeta}
\vth^{\sscr EH}|_{\p\Si}+\vth^{\sscr EH}|_{\p{\cal T}^\pm} = \mp2\d\b \eps_S,
\ee
where $\sinh\beta$ is the scalar product between the time-like and space-like outgoing normals, see Fig.~\ref{Fig1} and Appendix~\ref{AppA} for further relations. 
We consider first the case of orthogonal corners, $\b=0$. 
The corner variation vanishes, and we are left with the 3d bulk terms only. 
These vanish  if $\d\tl\Pi^{\m\n}\pb{\p M}0$,  as can be seen from the second equality of \eqref{thEH}: The Einstein-Hilbert action has thus a well-defined variational principle if we fix the momentum on the boundary, namely if we use Neumann boundary conditions.\footnote{This is true only in 4 dimensions, which is the only case in which the second equality of \eqref{thEH} holds. See e.g. \cite{Krishnan:2016mcj} for the Neumann boundary term in other dimensions.}
 
If instead we want to use Dirichlet boundary conditions, namely fix the induced metric so that
$\d q_{\m\n}\pb{\p M}0$, we see from the first equality of  \eqref{thEH} that there is a left-over term. To cancel it, we need to add the Gibbons-Hawking-York boundary Lagrangian
\be
\ell^{\sscr GHY} = 2sK \eps_\Si.
\ee
The  Lagrangian with a well-defined Dirichlet variational principle is thus
\be
L^{\sscr GHY} = L^{\sscr EH}+d\ell^{\sscr GHY}.
\ee

These two options are probably the ones most commonly considered.\footnote{Alternatively, one may wish to work with a field space in which variations are left arbitrary, and deduce the appropriate variational principle for each pair of bulk-boundary Lagrangian interpreting the boundary variation as an equation of motion. For example, $L^{\sscr GHY}$ induces the homogeneous Neumann boundary condition $\tl\Pi^{\m\n}=0$, whereas the EH Lagrangian would lead to the inadmissible degenerate condition $q_{\m\n}=0$. This approach, emphasized and studied in \cite{Margalef-Bentabol:2020teu}, will not be pursued here.} However as pointed out in \cite{Anderson:2006lqb,Anderson:2010ph}, a better choice of boundary conditions is York's mixed boundary conditions \cite{Choquet-Bruhat:1980ysa,York:1986lje}, which hold fixed the conformal metric $\hat q_{\m\n}:=q^{-1/3}q_{\m\n}$ and $K$, because they lead to a better posed initial boundary value problem (See also \cite{Witten:2018lgb,Wieland:2018ymr,Hilditch2013,An:2021fcq}). 
With this choice, one needs to add a boundary Lagrangian given by \cite{York:1986lje}
\be
\ell^{\sscr Y} = \f23sK \eps_\Si.
\ee
This can be easily proved observing that
\be\label{mixedPQ}
s\left( K_{\m\n} \d q^{\m\n}-2\d K \right) \sqrt{q}
= -s\tl P^{\m\n}
\d \hat q_{\m\n}  - s\tl P_K \d K -s\f23\d(K\sqrt{q}),
\ee
where
\be
\tl P^{\m\n} := q^{1/3}(\tl\Pi^{\m\n}-\f13q^{\m\n}\tl\Pi),
\qquad
\tl P_K=\frac{4}{3}\sqrt{q}
\ee
are the five plus one momenta conjugated respectively to the conformal metric and extrinsic curvature trace.\footnote{The general $n$-dimensional version of this formula is
\[
\left( K_{\m\n} \d q^{\m\n}-2\d K \right) \sqrt{q}
= - q^{\f1{n-1}}\left(\tl\Pi^{\m\n}-\f1{n-1}q^{\m\n}\tl\Pi\right)\d \hat q_{\m\n}  - 2\f{n-2}{n-1} \sqrt{q}\d K - \f2{n-1}\d(K\sqrt{q}).
\] For $n=3$, the required boundary term is thus $K\sqrt{q}$, namely one-half the GHY term. This is the numerical coefficient deduced in \cite{Detournay:2014fva} using asymptotic fall-off conditions for AdS and flat spacetimes. The nature of such asymptotic conditions was left as an open question there. The equation above suggests that they were of York's type.  However, since the Neumann boundary term in $n$ dimensions \cite{Krishnan:2016mcj} is given by $(4-n)K$, for $n=3$ it also gives one half of the GHY term. We think that in this case York's boundary conditions can be disregarded as ill-defined since, according to the uniformization theorem, the 2d conformal metric does not contain enough information about the boundary. Hence, the one-half GHY term should be considered of Neumann type.}

York's ``mostly-Dirichlet" mixed boundary conditions include the conformally flat initial data often used in numerical relativity \cite{Gourgoulhon2007}.
A peculiarity of this choice however is that the momenta don't commute, since 
next to the canonical pair $(\tl P_K,K)$, the remaining five pairs satisfy
(omitting $\d^{(3)}(x,x')$ to shorten the expressions)
\be\label{comm}
\{\hat q_{\m\n},\hat q_{\r\s}\} = 0, \qquad \{\hat q_{\m\n},\tl P^{\r\s}\} = \d^{\r\s}_{(\m\n)} - \f13 q_{\m\n} q^{\r\s},\qquad 
\{\tl P^{\m\n},\tl P^{\r\s}\} = \frac 13 \left( \hat{q}^{\rho\sigma}\tl P^{\m\n}-\hat{q}^{\m\n}\tl P^{\rho\sigma}\right).
\ee
The non-commutativity also implies that the flipped option of taking ``mostly-Neumann" mixed boundary conditions with fixed traceless momentum and the induced metric determinant is not admissible.\footnote{Which is good, because 
\be\nn
\tilde{\Pi}^{\m\n} \d q_{\m\n}= -\hat q_{\m\n}\d \tl P^{\m\n} + \f23 \tl\Pi\, \d\ln\sqrt q
\ee
shows that the EH Lagrangian would lead to a well-defined variational principle in this case as well, breaking the expected injective relation between choice of boundary Lagrangian and choice of boundary conditions.}

The three options considered above can be simultaneously treated taking as boundary Lagrangian
\be\label{ellc}
\ell^b:= bs K \eps_\Si,
\ee
with $b=2,2/3,0$ respectively for Dirichlet, mixed and Neumann boundary conditions.

In the following, it will be convenient to use different notations for the space-like and time-like boundaries. In doing so, we keep the notation $(n,u)$ for the space-like quantities, and introduce bars to distinguish the time-like or null boundaries, $(\bar n, \bar u)$. These notations are the ones summarized in Fig.~\ref{Fig1} and Table~\ref{tab:defn}. Accordingly, the action principle with orthogonal corners reads
\be\label{traceK}
S=\int_M L^{\sscr EH}+d\ell^b = \int_M R\eps -b\int^{\Si_1}_{\Si_0} K \eps_\Si +b\int_{\cal T}\KB\eps_{\cal T}. 
\ee

%--------------------------------------------------------------
\subsection{Non-orthogonal corners}
%--------------------------------------------------------------

If one allows for non-orthogonal corners, additional 2d boundary terms are potentially needed in the action principle, to compensate for variations like \eqref{dbeta}. The precise form of these variations depends on the type of corner considered, see \cite{Lehner:2016vdi} for a comprehensive analysis. We restrict attention here to the corner between a space-like and a time-like boundary, as in Fig.~\ref{Fig1}.
To cancel the corner variation \eqref{dbeta} we have two options: either we fix $\b$ so that $\d\b=0$, or we fix the induced metric so that $\d\eps_S=0$. In the first case, no boundary term is needed. In the second case, we need to add the Hayward boundary term
\be\label{ellH}
\ell^{\sscr H} = 2 \beta \eps_S.
\ee

The second option is consistent with Dirichlet boundary conditions, since fixing the induced metric $q$ also fixes $\eps_S$. This is indeed how Dirichlet boundary conditions with non-orthogonal corners are usually treated, see e.g. \cite{Hayward:1993my}. 
Fixing $\beta$ instead has the flavour of a Neumann-type condition, since it is easy to see that $\beta$ captures a metric component that is not part of the induced metric. But in fact, it is also not part of the momentum, but a combination of lapse and shift instead, see \eqref{betag}. Hence, it is an \emph{additional} condition to be provided.
This additional condition is a priori not needed for the well-posedness of the initial value boundary problem. In fact, the corner contribution can always be thought of as part of the space-like hypersurface, and then its variation corresponds simply to a change in the state, and not of the boundary conditions \cite{Harlow:2019yfa}. 
Furthermore, it is a change of state associated to a different choice of lapse and shift, and which can thus be considered irrelevant to characterize different physical solutions.\footnote{We thank Michael Anderson and Zhongshan An for clarifying discussions on this issue.}
 On the other hand, if the finite boundary is considered as a part of the characterization of the observer, then a solution with different $\b$ would be on the same status as, say, a Kerr solution with different values of the asymptotic lapse and shift, namely corresponding to boosted or rotated black holes. Hence, within the context of thinking of the gauge degrees of freedom broken by the boundary as physical, it is of interest to consider $\b$ as part of the phase space. Indeed, this choice will be justified by the canonical analysis of the boundary terms done in Section~\ref{SecCan}.
 
 A similar logic can be applied to the case of York's mixed boundary conditions. Since they leave the determinant of the induced metric free, it seems reasonable to us to take $\d\b=0$ also in this case, even though it is not required by the well-posedness of the initial value problem \cite{Anderson:2006lqb,An:2021fcq}. Again, this will be justified by the canonical analysis reported below. 
These choices are summarized in Table~\ref{Tablebc}.

As before, we can treat all cases with a generic corner Lagrangian
\be\label{ellcorner}
\ell^{c} = c \beta \eps_S,
\ee
with $c$ needing to be $2,0$ and $0$ respectively for Dirichlet, mixed and Neumann boundary conditions.
The action principle with non-orthogonal corners thus reads
\begin{align}\label{traceK2d}
S=\int R\eps - b\int^{\Si_1}_{\Si_0} K\eps_\Si+ b \int_{\cal T} \KB\eps_{\cal T} + c\int^{S_1}_{S_0} \beta\, \eps_S.
\end{align}
For Dirichlet boundary conditions, \eqref{traceK} and \eqref{traceK2d} are referred to as trace-K actions in the literature. Accordingly, we will refer to them as $b$-generalized trace-K actions. 

\begin{table}[h] \begin{center}
  \begin{tabular}{l|c|c|c|c}  
\emph{boundary conditions} & \emph{quantity fixed on boundary} & \emph{value of $b$} & \emph{quantity fixed at corner} & \emph{value of $c$} \\ \hline
Dirichlet & $q_{\m\n}$ & 2 & $\g$ & 2\\
York & $(\hat q_{\m\n}, K)$ & $2/3$ & $\b$ & 0\\
Neumann & $\tl\Pi^{\m\n}$ & 0 & $\beta$ & 0
\end{tabular} \end{center}   
\caption{\label{Tablebc} \emph{\small{Different boundary conditions and their boundary and corner Lagrangians.}}}\end{table}

%--------------------------------------------------------------
\section{Surface charges from covariant phase space}
%--------------------------------------------------------------

The first method we are going to use to compute the charges associated with the different boundary conditions is the covariant phase space, in particular with the prescription of \cite{Freidel:2020xyx}, which we briefly review here.
We use the notation of \cite{Harlow:2019yfa,Freidel:2021cbc}, where $\d$ and $I$ are respectively the exterior derivative and internal product in field space. We denote $\th^\ssL$ the (integrand of the) symplectic potential associated to a bulk Lagrangian $L$, which satisfies $\d L\approx d\th^\ssL$.\footnote{If the latter equality is taken as the definition of the symplectic potential, it makes it ambiguous by the freedom to add any exact 3-form. We fix this ambiguity by taking always the `bare' choice, which can be justified defining the symplectic potential  by using the Anderson homotopy operator prescription, see \cite{Anderson,Compere:2018aar,Freidel:2020xyx}.} The quantities
\be
\Th^\ssL:=\int_\Si \th^\ssL, \qquad \Om^\ssL=\int_\Si \om^\ssL, \qquad \om^\ssL=\d\th^\ssL,
\ee
define respectively a (pre-)symplectic potential and (pre-)symplectic 2-form in the space of fields associated with the hypersurface $\Si$ and the Lagrangian $L$. The formalism can be used to compute canonical generators of the infinitesimal symmetries $\d_\xi$ of $L$, by seeking functionals $H^\ssL_\xi$ that would satisfy  $\d H^\ssL_\xi=-I_\xi\Om^\ssL$.\footnote{We use throughout $I_\xi$ as short-hand notation for $I_{\d_\xi}$.}
For field-independent diffeomorphisms and in the absence of anomalies, one has
\cite{Iyer:1994ys,Wald:1999wa}
\be\label{IxiOm}
-I_\xi\Om^\ssL = \int_S \d q^\ssL_\xi-i_\xi\th^\ssL,
\ee
where
$q^\ssL_\xi$
is the Noether charge associated with the conserved Noether current
\be\label{jN}
j^\ssL_\xi:=I_\xi\th^\ssL-i_\xi L \approx d q^\ssL_\xi, \qquad dj^\ssL_\xi\approx 0.
\ee
For diffeos that are tangent to the corner,  $\xi\in TS$, the second term in \eqref{IxiOm} vanishes, and the canonical generator can be identified with the Noether charge, as customary from the application of Noether's theorem on flat spacetime. For $\xi\notin TS$, this is no longer the case. One way to understand this is that these diffeomorphisms move the corner, and by doing so, one is sensitive to degrees of freedom that could be entering or escaping the causal domain of $\Si$. One is thus effectively dealing with an open system, and the construction of canonical generators is more subtle. This is an active field of research (see e.g. \cite{Barnich:2011mi,Flanagan:2015pxa,Compere:2020lrt,Chandrasekaran:2020wwn,Freidel:2021cbc,Wieland:2020gno} and references therein), particularly relevant to the study of gravitational radiation and with applications to entanglement and quantum gravity \cite{DonnellyFreidel,Freidel:2020ayo,Geiller2017}.

One situation with a simple solution to this problem is in the presence of conservative boundary conditions, that freeze the number of degrees of freedom available. This effectively closes the system and makes a Hamiltonian available, as we review next.

\subsection{Hamiltonians from conservative boundary conditions}

As explained in \cite{Wald:1999wa} and more explicitly developed in \cite{Harlow:2019yfa} and especially \cite{Freidel:2020xyx}, Hamiltonians can be constructed if we restrict the variations to preserve some given boundary conditions along $\cal T$. Consider the boundary Lagrangian $\ell$ required by a well-defined variational principle with given boundary conditions at $\cal T$. By construction, restricting the variations to those preserving the boundary conditions, the boundary variation of $L$ must be equal and opposite to the variation of $\ell$, up at most to a corner term: $\th^\ssL\pb{\cal T}-\d\ell+d\vth$. 
As a consequence, it is possible to redefine the symplectic potential and 2-form as
\be\label{thom}
\th:=  \th^{\sscr L} +\d\ell-d\vth, \qquad \om=\d\th = \d\th^\ssL -d\d\vth.
\ee
 In this way, one automatically has vanishing symplectic flux across the boundary, when the boundary conditions are imposed:
\be\label{bc}
\th|_{\rm b.c.}\pb{\cal T}0 \quad \Leftrightarrow\quad \th^{\sscr L}|_{\rm b.c.}\pb{\cal T}-\d\ell+d\vth \qquad \Rightarrow \qquad \om|_{\rm b.c.}\pb{\cal T}0.
\ee
This condition guarantees that the system is closed, and the defining equation \eqref{IxiOm} integrable to yield a Hamiltonian generator.\footnote{The last equation in \eqref{bc} is Wald's sufficient condition for integrability. It can also be derived requiring the condition that an Hamiltonian vector field (in field space) preserves the symplectic form, $\d_\xi \om = \d(I_\xi\om) = 0$, which implies $I_\xi\om = \d h_\xi$,
and if there are no anomalies, we also have $\d_\xi \om = \pounds_\xi \om \approx d(i_\xi\om)$.
}
The new quantities in \eqref{thom} depend on the pair $(L,\ell)$ of bulk-boundary Lagrangians, dependence which we don't make explicit in order to keep the notation light.
The resulting Hamiltonian charge is\footnote{This can be seen evaluating
\begin{align}\nn
-I_\xi \om &= -I_\xi\d\th = -I_\xi\d\th^{\sscr L} + dI_\xi\d\vth \approx d( \d q_\xi - i_\xi\th^{\sscr L} - q_{\d\xi}+ i_\xi d\vth- \d I_\xi\vth  + I_{\d\xi}\vth), 
\end{align}
where we used
\be\nn
I_\xi\d\vth = \d_\xi \vth-\d I_\xi\vth = \pounds_\xi\vth+I_{\d\xi}\vth - \d I_\xi\vth = d(i_\xi\vth)+i_\xi d\vth  - \d I_\xi\vth +I_{\d\xi}\vth.
\ee
For field-independent diffeomorphisms and on-shell of \eqref{bc} we obtain \eqref{HamC}. Further details can be found in e.g. \cite{Harlow:2019yfa,Freidel:2021cbc}.
Here we assumed that no anomalies are present, but the resulting formula \eqref{HamC} is valid also in the anomalous case. This was proved in \cite{Freidel:2021cbc}, and one has to use the fact that boundary conditions are consistent, namely $\D_\xi\th|_{\rm b.c.}\pb{\cal T}0$.}

\be\label{HamC}
-I_\xi\Om =\d H_\xi, \qquad H_\xi =\int_S q_\xi+i_\xi\ell-I_\xi\vth,
\ee
up to a constant of integration in field space that we will come back to below in Section~\ref{SecSub}.
We also notice that this expression \emph{always} coincides with the Noether charge, since
\be
j_\xi = I_\xi\th -i_\xi (L+d\ell) = I_\xi\th^\ssL +\d_\xi\ell-I_\xi d \vth-i_\xi L -i_\xi d\ell =d (q_\xi^\ssL+ i_\xi\ell -I_\xi\vth).
\ee
Thanks to the conservative boundary conditions, all allowed diffeomorphisms acquire the status of the tangential diffeomorphisms of the original construction: they are integrable transformations, with Hamiltonian generator given by the Noether charge. The key novelty is that the Noether charge is associated not only to $L$ as in the original prescription of \cite{Iyer:1994ys}, but with the pair $(L,\ell)$ through the formula \eqref{HamC}. It appeared in this form in the work \cite{Freidel:2020xyx} by Freidel, Geiller and Pranzetti, and accordingly we will refer to it as FGP prescription.

In this formula, the allowed diffeomorphisms are those that preserve the boundary conditions whose imposition makes the charges integrable.
This means that $(i)$ they cannot move the boundary $\cal T$, namely
\be\label{xitang}
\xi\in T{\cal T}\quad \Leftrightarrow \quad \xi\cdot \bar n=0,
\ee
which equals $\xi\cdot u=0$ in the case of orthogonal corners, and $(ii)$, $\d_\xi F(g_{\m\n})=0$, where $F$ are the boundary conditions chosen.

%--------------------------------------------------------------
\subsection{Corner symplectic potential}
%--------------------------------------------------------------

The only aspect of the prescription \eqref{HamC} that requires some care is the determination of the corner symplectic potential. It can be in principle  computed using Anderson's homotopy operator as argued in \cite{Freidel:2020xyx}, but in practise it is simpler to derive it taking the variation of the boundary Lagrangian, and arranging it in such a way that the boundary field equations are consistent with the boundary conditions one is imposing. 
Let us see explicitly this strategy at play with $\ell^b$. To compute its variation, we  use the standard result
\be\label{deltaK}
\delta K= -\frac 12 K^{\m\n}\delta g_{\m\n} + g^{\r[\s}n^{\m]} \na_\m \delta g_{\r\s} +\f12q_\m^\rho \na_\rho\left(q^{\m}_\n\d n^\n\right).
\ee
The second term is proportional the symplectic potential $\th^{\sscr EH}$, see \eqref{thg}, and the third term can be written in terms of 
the induced covariant derivative on the hypersurface  $D_\m$.\footnote{Here we used
\begin{align}\nn
& \int_\Si D_\m (q^\m_\n \d n^\n) d\Si = \int_{\p\Si} q^\m_\n \d n^\n dS_\m = -s \int_{\p\Si} u_\m \d n^\m dS
\qquad dS_\m=-su_\m dS.
\end{align}}
Applying this formula to \eqref{ellc}, we find
\begin{align}
\d \ell^b &=\f b2 s\left(\left(K q^{\mu \nu}-K^{\mu \nu}\right) \delta g_{\mu \nu}+
2g^{\alpha [\beta} n^{\lambda]} \nabla_{\lambda} \delta g_{\alpha \beta} 
+ D_{\mu}\left(q^{\mu}_{\nu}\d n^\n \right)\right) \eps_\Sigma \nn\\
& = \f b2\left(-\th^{\sscr EH} + s(K_{\m\n}-q_{\m\n}K)\d q^{\m\n} \eps_\Sigma+ d\vth^{\sscr EH} \right), \label{dell}
\end{align}

To determine the corner symplectic potential of $\ell^b$, the bulk term must be consistent with the boundary conditions we want to impose, as to reproduce (the first of) \eqref{bc}. Rearranging the terms in \eqref{dell}, we find
\begin{align}
\th^{\sscr EH}+\d\ell^b &= (1-\f b2)\th +\f b2s(K_{\m\n}-q_{\m\n}K)\d q^{\m\n}  \eps_\Si + \f b2 d\vth^{\sscr EH} \label{thell} \nn\\
& = s\left( (K_{\m\n}- \f b2 q_{\m\n}K)\d q^{\m\n} +(b-2)\d K \right)  \eps_\Si+d\vth^{\sscr EH}.
\end{align}
We can explicitly check that the term in bracket in the second equality vanishes accordingly to the boundary conditions chosen: $\d q_{\m\n}\pb{\cal T}0$ for $b=2$, $\d\tl\Pi^{\m\n}\pb{\cal T}0$ for $b=0$ (see \eqref{thEH}), and $\d\hat q_{\m\n}\pb{\cal T}0\pb{\cal T}\d K$ for $b=2/3$ (see \eqref{mixedPQ}). Therefore, we conclude that the corner symplectic potential of $\ell^b$ is precisely $\vth^{\sscr EH}$, irrespectively of these values of $b$. 

The fact that this construction yields a consistent non-vanishing $\vth$ for $b=0$ is quite remarkable. It leads to the suggestion of taking a non-vanishing corner symplectic potential also for Neumann boundary conditions, even if the boundary Lagrangian is zero in this case. One may discard this possibility, but as we will see below, keeping it allows one to introduce an integrable charge for the Einstein-Hilbert action valid also for non-tangential diffeos, and which reduces to the Komar expression in the case of isometries.

%--------------------------------------------------------------
\subsection{Charges for Dirichlet, mixed and Neumann boundary conditions}
%--------------------------------------------------------------

We now apply the prescription \eqref{HamC} to the Lagrangian $L^b:=L^{\sscr EH} + d\ell^b$. This requires evaluating the three terms in \eqref{HamC} and their pull-backs on the corner $S$ intersection of $\Si$ and $\cal T$. We consider the corner in the future of $\cal T$, so that the outgoing time-like normal is future-pointing. Our conventions for the volume forms, orientations and pull-backs are reported in Appendix~\ref{AppA}.
The symplectic potential of $L^{\sscr EH}$ is given in \eqref{thEH}, and its Noether charge is the Komar 2-form \cite{Iyer:1994ys}
\be\label{Komar}
q^{\sscr EH}_{\xi}=-\f12\eps_{\m\n\r\s}\na^\r\xi^\s dx^\r\w dx^\s \pb{S}2 n_\m u_\n \na^{[\m}\xi^{\n]}\eps_S.
\ee
Next, the pull-back of $\ell^b$ on the time-like boundary $\cal T$ gives
\begin{align}& i_\xi \ell^b \pb{\cal T} b\KB i_\xi \eps_{\cal T} = \f b2\KB u^\m \xi^\n \eps_{\m\n\r\s}dx^\r\w dx^\s 
\pb{S} b\KB \xi\cdot n\, \eps_S
\end{align}
The last ingredient is the corner symplectic potential just evaluated. Its pull-back on the future boundary of $\cal T$ gives
\be
\vth = \vth^{\sscr EH} \pb{\p{\cal T}^+} -u_\m\d n^\m (-\eps_S)= n_\m u_\n \d g^{\m\n} \eps_S, 
\qquad I_\xi \vth = -2 n_\m u_\n \na^{(\m}\xi^{\n)} \eps_S.
\ee
Adding up the three terms, the Hamiltonian charge is found to be
\begin{align}\nn
H_\xi^b&=\int_S q^{\sscr EH}_{\xi}+i_\xi\ell^b-I_\xi\vth^b = \int_S 2n_\m u_\n ( \na^{[\m}\xi^{\n]} + \f b2 \KB \xi^\m u^\n + \na^{(\m}\xi^{\n)} )  \eps_S 
\nn\\ &= 2\int_S n_\m u_\n ( \na^{\m}\xi^{\n} + \f b2 \KB \xi^\m u^\n )  \eps_S 
\nn\\ &=-2 \int_S n^\m \xi^\n (\KB_{\m\n} - \f b2 \qB_{\m\n}\KB )  \eps_S, \label{charge}
\end{align}
where in the last step we used that for orthogonal corners we can take $u\equiv \bar n$, therefore $\xi\cdot u=0$,  $n^\m=\bar q{}^{\m\n} n_\n$ and $n^\m\na_\m u_\n=n^\m \bar K_{\m\n}$. 

From this general formula we can read the three special cases we have been discussing so far.
For Dirichlet boundary conditions, $b=2$, we have
\begin{align}
H_\xi^{\sscr BY}&=-2 \int_S n^\m \xi^\n (\KB_{\m\n} - \qB_{\m\n}\KB )  \eps_S 
= -2 \int_S n^\m \xi^\n \bar \Pi_{\m\n}  \eps_S. \label{chargeBY}
\end{align}
This is the result of \cite{Harlow:2019yfa} (see also \cite{Wald:1999wa,Freidel:2020xyx}): the Hamiltonian generating the boundary symmetries in the covariant phase space is the Brown-York surface charge. 
For York's mixed boundary conditions, we find 
\begin{align}
H_\xi^{\sscr Y} = 
-2 \int_S n^\m \xi^\n (\KB_{\m\n} - \f 13 \qB_{\m\n}\KB )  \eps_S
= -2 \int_S n^\m \xi^\n \bar \Pi_{\mean{\m\n}}  \eps_S,
\end{align}
namely the surface charge is the traceless part of the ADM momentum on the time-like boundary. 
This result appeared recently in \cite{An:2021fcq}.
Finally for Neumann boundary conditions, the trace part of the extrinsic curvature drops out and we are left with
\begin{align}\label{HN}
H_\xi^{\sscr N} = 
-2 \int_S n^\m \xi^\n \KB_{\m\n} \,  \eps_S
=-2 \int_S n^\m \xi^\n (\bar\Pi_{\m\n}-\f12\bar q_{\m\n}\bar\Pi)  \eps_S.
\end{align}

The last expression can be taken as definition of integrable charge for the Einstein-Hilbert action valid also for diffeomorphisms non-tangential to the corner $S$, for which the usual prescription fails, as remarked below \eqref{jN}. 
It is quite a non-trivial step, since with the usual prescription one obtains integrable charges only for tangential diffeomorphisms, and follows from taking seriously the FGP prescription and the construction \eqref{thell} of the corner symplectic potential.
The new Einstein-Hilbert charge 
reduces to the Komar expression for (arbitrary) tangential diffeomorphisms. 
This can be seen starting from the second line of \eqref{charge} with $b=0$, and using $n\cdot u=n\cdot \xi=u\cdot \xi=0$ to prove that the symmetrization in $n$ and $u$ vanishes.
We also notice that for non-tangential diffeomorphisms, \eqref{HN} reduces to the Komar expression in the case of isometries. 
This can be immediately seen again from the second line of \eqref{charge} with $b=0$ and using the Killing equation. 
Hence, \eqref{HN} does provide an extension of the Komar formula to non-isometric diffeomorphisms endowed with an interpretation of Hamiltonian generator for Neumann boundary conditions. 

The charges \eqref{charge} can be split into energy and angular momentum,  introducing a decomposition of the diffeomorphism as 
\be\label{xiN}
\xi^\m = N n^\m +N^\m, \qquad N\cdot n=0.
\ee
Notice that since we are already restricting the diffeos to satisfy \eqref{xitang}, $N^\m$ only has components tangent to the corner. Then, 
\be
n^\m N^\n \bar K_{\n\m} = n^\m N^\n \bar q_\n^\r\na_\r u_\m = - u^\m N^\n \bar q_\n^\r\na_\r n_\m =- u^\m N^\n \g_\n^\r\na_\r n_\m =
- u^\m N^\n K_{\n\m}. 
\ee
As for the piece proportional to lapse, it can be expressed in terms of the 2d extrinsic curvature 
$k=\bar K +n^\m n^\n\na_\n u_\n$ (see \eqref{littlek} in the Appendix, here adapted to orthogonal corners). Adding up, and using $\xi\cdot n=-N$, 
\eqref{charge} gives
\begin{align}\label{chargelapse}
H_\xi^b&= -2 \int_S \left(N\Big(k + \f{b-2}2 \KB \Big) - N^\m u^\n K_{\m\n} \right)\eps_S.
\end{align}
In Section~\ref{SecCan}, we will reproduce this expression using canonical methods.

The term proportional to $N$ is the energy, whereas the term proportional to $N^\m$ 
is the corner diffeomorphism charge and contains the angular momentum.
We see that changing the boundary conditions leaves the angular momentum invariant but changes the notion of energy of the system. 
The fact that integrable charges obtained through the imposition of boundary conditions depend on the latter was expected \cite{Iyer:1995kg}, and we are seeing here the results of a quantitative analysis. 
This dependence is after all understandable: in the open case there is no general notion of energy, so it makes sense that when we close it, the notion of energy depends on \emph{how} we close the system. 
See also \cite{Brown:1990fk} for earlier discussions on the relation between energy and boundary conditions in general relativity.

The quasi-local charges \eqref{charge} are defined on the surface corner $S$ of a finite region of spacetime. To study what happens for asymptotic charges, we need to first consider the required subtraction terms. We will do so in the next subsection, and in the next Section we will use the Kerr spacetime to explore the explicit quasi-local and asymptotic values of the charges and see how they are affected by the choice of boundary conditions.

%--------------------------------------------------------------
\subsection{Subtraction terms and symplectic renormalization}\label{SecSub}
%--------------------------------------------------------------
The quasi-local expressions \eqref{charge} or equivalently \eqref{chargelapse} are fine as quasi-local charges at finite distance, but they diverge when the corner is pushed to spatial infinity. This is a familiar result from the Brown-York analysis, and the standard procedure is to remove the divergence with a subtraction term depending on a background solution, typically Minkowski. From the covariant phase space perspective, this is a natural procedure that amounts to the simple fact that when integrating \eqref{HamC} one can take into account a non-vanishing constant of integration in field space \cite{Wald:1999wa}. This reference or background solution can be taken to be Minkowski, and produces the subtraction term of the Brown-York analysis leading to finite expressions as pointed out in \cite{Ashtekar:2008jw}.\footnote{There the finiteness was attributed to the use of tetrads in the first order formalism, but we believe the result applies to any formulation, and follows from the fact that by plugging in the fall-off condition on the variables in the variational formula \eqref{IxiOm}, the leading order Minkowski contribution is eliminated directly being a fixed background.} We now show that this procedure can be generalized to $b\neq 2$, and that it can also be understood in the framework of symplectic renormalization, which plays an important role for subtracting analoguous divergences at null infinity \cite{Compere:2018ylh,Compere:2020lrt,Freidel:2021yqe}. Namely instead of removing the divergence via a background solution, we can renormalize the charge using the prescription \eqref{HamC} and subtracting the contribution that would come from a boundary Lagrangian $\ell_{\sscr div}$ that captures the divergences of $L$, namely
\begin{align}
H^{{\rm R}}_\xi &= H_\xi -\int_S i_\xi\ell_{\sscr div} -I_\xi\vth_{\sscr div}. 
\end{align}

In the present context, we restrict attention to asymptotically flat metrics at spatial infinity, and the leading divergence of $H^b_\xi $ comes from the Minkowskian behaviour of the charge. Therefore, 
\be
\ell^b_{\sscr div}=b\bar K_\eta\, \eps_\Si, \qquad i_\xi\ell^b_{\sscr div}\pb{S} b\bar K_\eta \, \xi\cdot n\, \eps_S, \qquad \vth^b_{\sscr div}=0.
\ee
Then the renormalized charge is
\begin{align}
H^{{\rm R}b}_\xi &= H^b_\xi -\int_S i_\xi\ell^b_{\sscr div} \\\nn
&=-2 \int_S n^\m \xi^\n \Big(\KB_{\m\n} - \f b2 \qB_{\m\n}(\KB-\bar K_\eta) \Big)  \eps_S \\\nn
&=-2 \int_S \left(N\Big(k -\f b2k_\eta+ \f{b-2}2 \KB \Big) + N^\m n^\n \KB_{\m\n} \right)\eps_S,
\end{align}
where in the last step we used \eqref{xiN} and the fact that in Minkowski,
\be\label{keta}
\bar K_\eta=k_\eta=\f2r.
\ee

Restoring the $16\pi G$ factors, we define the $b-$generalized energy and angular momentum as follows,
\begin{align}\label{BYenergy}
E&=H^{{\rm R}b}_{n}= -\f1{8\pi G}\int (k-\f b2 k_0+ \f{b-2}2 \KB)\eps_S
\\
J&=H^b_{\p_\phi}= -\f1{8\pi G}\int n^\m \bar K_{\m\phi} \eps_S,
\end{align}
which correspond respectively to the generator of unit-lapse hypersurface-orthogonal diffeomorphisms, and rotations around a fiducial vertical axis fixed say by asymptotic flatness.
We see that the boundary conditions do not affect the angular momentum, as expected since this charge does not see the symplectic flux, and furthermore is independent of renormalization since it coincides with its quasi-local value. The energy on the other hand depends explicitly on both the boundary conditions and the renormalization.

%--------------------------------------------------------------
\subsection{Residual diffeomorphisms}
%--------------------------------------------------------------

Let us give a few more details about the allowed diffeomorphisms. As discussed above, they must preserve the boundary as well as the boundary conditions. Preserving the boundary means $\xi\cdot \bar n=0$, so that $\xi\in T\cal T$. 

The set of diffeomorphisms that respect Dirichlet boundary conditions satisfy
\be\label{bKV}
\bar q_\m^\rho \bar q_\n^\sigma \d_\xi g_{\rho\sigma} |_\mathcal{T}= 0 \quad \Leftrightarrow \quad \bar D_{\left(\m\right.}\xi_{\left.\n\right)}=0.
\ee
These are boundary Killing vectors. These are not required to be isometries of the whole spacetime, so this condition does not restrict the bulk metric. 

In the case of York's mixed boundary conditions,  fixing the conformal metric is preserved by conformal Killing vectors of the boundary,
\be\label{confbKV}
(\bar q_\m^\r \bar q_\n^\s-\f13\bar q_{\m\n}\bar q^{\r\s}) \d_\xi g_{\rho\sigma} \stackrel{\mathcal{T}}= 0 \quad \Leftrightarrow \quad
\bar D_{\left(\m\right.}\xi_{\left.\n\right)} - \frac 13 \bar q_{\m\n}\bar D\cdot \xi\stackrel{\mathcal{T}}=0.
\ee
While the condition on the trace of the extrinsic curvature constrains the projection of $\xi$ along the acceleration. To see this, we start from the variation \eqref{deltaK}. Specializing to a diffeomorphism, it can be written in the following form, 
\be
\d_\xi K= -2(K^{\m\n}-\f s2 K n^\m n^\n) \na_\m\xi_\n - n^\m(g_{\m\n}\square+R_{\m\n})\xi^\n + s n^\m n^\n n^\r  \na_\m \na_\n \xi_\r,
\ee
which makes it conveniently manifest that it would vanish exactly for a Killing vector. Using the orthogonality and conformal boundary Killing properties of $\xi$, it reduces to
\be
\d_\xi K=  -\f23 K D\cdot\xi -sK a\cdot\xi - R_{\m\n}n^\m\xi^\n.
\ee
Imposing this to be zero (and restoring the bars and $s=1$ appropriate to the time-like boundary we are interested in), we obtain 
\be
\f23 \bar K \bar D\cdot\xi = -\bar K \bar a\cdot\xi - R_{\m\n}\bar n^\m\xi^\n.
\ee
This scalar equation between components of $\xi$ completes the restriction given by \eqref{confbKV}.
Unless the right-hand side vanishes, the residual diffeomorphisms are not the same as in the Dirichlet case.

One can proceed similarly to establish the residual diffeomorphisms in the case of Neumann boundary conditions.

%--------------------------------------------------------------
\section{Kerr example}
%--------------------------------------------------------------
In this section we restore the $16\pi G$ factors. 
To get some further intuition about the meaning of these different charges, we consider their explicit values in the case of the Kerr solution.
Integrating the Komar form \eqref{Komar} on a 2-sphere at constant $(t,r)$ for the two Killing vectors $\p_t$ and $\p_\phi$ one gets
\be\label{KK}
Q^{\sscr EH}_{\p_t}=  \f M{2G}, \qquad Q^{\sscr EH}_{\p_\phi}=  - \f{Ma}G.
\ee
This result is independent of $r$ since the Noether current $j^{\sscr EH}_\xi$ vanishes in vacuum for a Killing vector.  
This is the standard Noether charge for the Einstein-Hilbert Lagrangian. We now compute the Hamiltonian charge \eqref{HamC} associated with different boundary conditions. First of all, we observe that $I_\xi\vth$ is proportional to the Killing equation and thus vanishes for both $\p_t$ and $\p_\phi$.  Since $\p_\phi$ is tangential to the corner,   $ i_{\p_\phi}\ell^b=0$, and the Hamiltonian charge coincides with the Komar expression, 
$H_{\p_\phi}=-Ma/G$.  However,  this is not the case for the charge generated by $\p_t$. Using $n_t:=n\cdot\p_t=-(-g^{tt})^{-1/2}$, we evaluate
\be
\int_{S^2} i_{\p_t}\ell^b = \f b{16\pi G} \int_{S^2} \bar Kn_t \eps_S
=-\frac{b}{4 G} \left(r-M+\f{\D}a \arctanh\frac{a}{r}\right),
\ee
where $\D:=r^2-2 Mr+a^2$.
Adding this up to the Komar expression according to \eqref{HamC},  we find
\begin{align}\label{bareExp}
H_{\p_t}&=
\f{b+2}{4G}M-\f b{4G}\left(r +\f{\D}a \arctanh\frac{a}{r}\right) 
\\\nn&=-\f b{2G}r+ \f{3b+2}{4G}M -\f b{6G}\f{a^2}r+O(r^{-2}).
\end{align}
This expression diverges linearly, as discussed above. 
We also notice that for $b=0$ the expression coincides with the value of the Komar charge alone \eqref{KK}, as to be expected from the equivalence in the case of isometries of the Neumann charge discussed below \eqref{HN}.
Adding the subtraction term, we arrive at the renormalized charge
\be\label{Hren}
H^{\rm R}_{\p_t}= H_{\p_t} - \f b{2G} r n_t = \f{b+2}{4G}M -\f b{4Gr}\left(M^2+\f{2a^2}{3}\right)+O(r^{-2}),
\ee
where we used
\be
n_t=-1+\f Mr+\f{M^2}{2r^2}+O(r^{-3}).
\ee

We see that the asymptotic value of the renormalized energy still depends on the choice of boundary conditions. 
For $b=2$ we recover the usual energy of the Kerr spacetime, namely $M$.
For mixed boundary conditions the asymptotic energy reduces to $2/3M$, and for Neumann boundary conditions to $M/2$.

We also remark that for all values of $b$, the quasi-local charge for the time-diffeomorphisms is $r$-dependent, a result familiar from the Brown-York papers. In this derivation based on \eqref{HamC}, the $r$-dependence is introduced by the contribution of the boundary Lagrangian, and 
captures the fact that 
the full quasi-local charge does not descend from an on-shell vanishing current as the Komar term alone.

In this Kerr example we found it natural to evaluate the energy using the (not hypersurface-orthogonal) Killing vector $\p_t$, but a more general choice for the energy is to take the hypersurface orthogonal time-like vector $Nn$, which can always be introduced. For Kerr, these two choices asymptotically align and require the same subtraction term. The resulting value of the energy is also very similar: the difference turns out to appear only at order $O(r^{-3})$. To eliminate all reference to the choice of diffeomorphism, we can also use the definition  \eqref{BYenergy} of generalized BY energy, which correspond to the generator of unit-lapse hypersurface-orthogonal diffeomorphisms.
The resulting expression is slightly more involved than \eqref{Hren},\footnote{This can be understood because the choice of a Killing vector sets to zero the $I_\xi\vth$ contribution to the charge, whereas \eqref{BYenergy} sees this contribution as well.} 
and with the help of Mathematica we find
\begin{align} \label{Eba}
E &=-\frac{r}{4G}\sqrt{\frac{\D}{a^2+r^2}}\left(  b-2 -2 b \sqrt{\frac{a^2+r^2}{\D}} 
+\frac{br (r-M) }{\D}\frac{\sqrt{a^2+r^2}}a\arctanh\frac{a}{\sqrt{a^2+r^2}}\right.\\\nn
&\qquad \left.+\frac{\sqrt{2} \left(r\D -M(r^2-a^2)\right) }{a \D\sqrt{M  \left(r\D+2M(2r^2+a^2)\right)}}\arctanh  \sqrt{\frac{2Ma^2}{ a^2(2 M+r)+r^3}} \right) 
\\ \nn &\stackrel{b=2 }{=}  -\frac{r}{4G}\sqrt{\frac{\D}{a^2+r^2}}\left(  \frac{2r (r-M) }{\D}\frac{\sqrt{a^2+r^2}}a\arctanh\frac{a}{\sqrt{a^2+r^2}}\right.
\\\nn
&\qquad \left.+\frac{\sqrt{2} \left(r\D -M(r^2-a^2)\right) }{a \D\sqrt{M  \left(r\D+2M(2r^2+a^2)\right)}}\arctanh  \sqrt{\frac{2Ma^2}{ a^2(2 M+r)+r^3}}-4 \sqrt{\frac{a^2+r^2}{\D}}\right) \\\nn
&\stackrel{a=0}{=}\f rG\left(1-\sqrt{1-\f{2M}r}\right) = \f MG-\f{M^2}{2Gr}+O(r^{-2}).
\end{align}
The expression with $b=2$ and $a=0$ can be recognized as the familiar BY result for Schwarzschild. Expanding the general expression \eqref{Eba} at spatial infinity, we find
\be
E =  \f{b+2}{4G}M -\f 1{6Gr}\left(3 M^2 +b a^2 \right)+O(r^{-2})
\ee
The different choice of diffeomorphism is reflected by the different subleading terms, but the asymptotic value is the same as \eqref{Hren}, in particular the dependence on the boundary conditions is the one already discussed.

%--------------------------------------------------------------
\section{Covariant surface charges with non-orthogonal corners}\label{SecNO}
\begin{figure}[h!]
\centering 
\includegraphics[width=4cm]{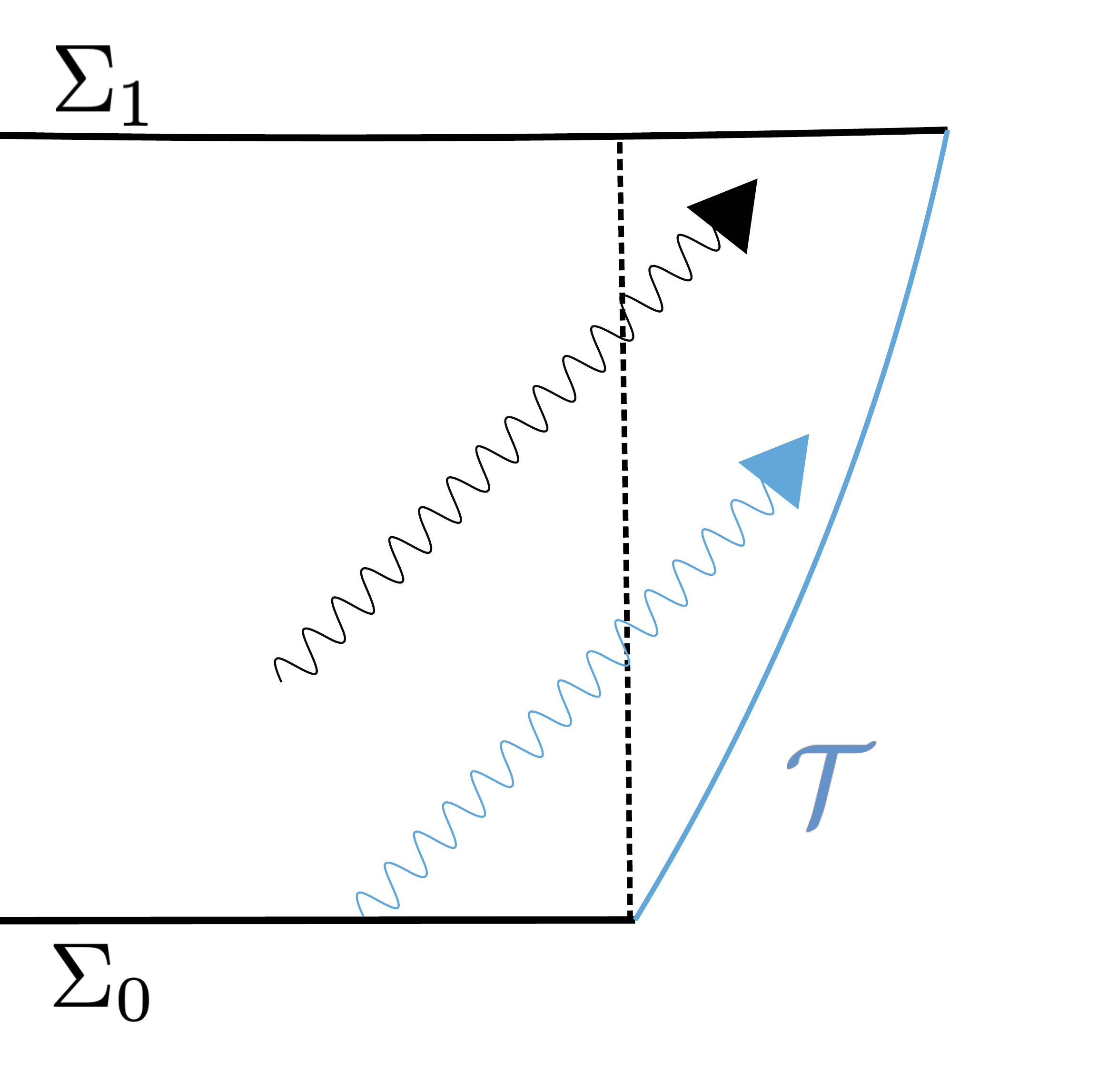}
\caption{ \label{Fig2}
\small{\emph{If the time-like hypersurface at which the boundary conditions are imposed is tilted with respect to the time-like evolution of the corner of $\Si$, symplectic flux can a priori leak through it, as in the top (and darker) arrow. In that case, the procedure to get integrable charges applies only to the `barred observers', namely those at rest with respect to a foliation of $\cal T$.} }}
\end{figure}

%--------------------------------------------------------------
In this Section we look at covariant phase space charge in the case of non-orthogonal corners, $\b\neq 0$. We can distinguish two classes of observers, those at rest with respect to the space-like foliation $\Si$, and those at rest along the time-like boundary. We may refer to them as respectively unbarred and barred observers, as in \cite{Brown:2000dz}. 
At the corner, these are related by the boost transformation with rapidity $\b$, see Fig.~\ref{Fig1} and \eqref{SO11}. 
Canonical methods, who fail to take into account the presence of leakage at the time-like boundary, can be used to compute charges for either class of observes \cite{Brown:2000dz}. However this may not be the case for covariant phase space methods. As reviewed earlier, the condition for the integrability of the covariant phase space charges requires a vanishing symplectic flux through the time-like boundary. Imposing boundary conditions at a non-orthogonal time-like boundary means that symplectic flux can a priori leak though the time-like evolution of an unbarred observer at the corner (for instance through late time null trajectories or time-like trajectories), see Fig.~\ref{Fig2}. 
Therefore, we will only construct covariant phase space charges for the barred observers. 

Apart from this conceptual difference, there is also a technical mathematical question raised by non-orthogonal corners. 
The formula \eqref{HamC} can be applied straightforwardly, however which boundary Lagrangian and corner symplectic potential should we use? As shown by \eqref{traceK2d}, the presence of non-orthogonal corners requires additional boundary terms in the action principle. How should these be taken into account in the prescription for the charges?

To address this question, we compute the charges using \eqref{HamC} with the same 3d boundary Lagrangian and corner symplectic potential as in the orthogonal case, simply ignoring any possible additional terms. 
We will then compare 
the result with the one obtained with canonical methods. The discrepancy will provide the answer to our question.
Notice that even using the same 3d boundary Lagrangian, a non-vanishing $\b$ shows up in the normals that appear in the formula.
In particular, we can pull-back at the corner using the barred basis $(\bar u,\bar n)$ instead of the unbarred basis $(n,u)$. The three ingredients for the charge are then
\begin{align}
& q^{\sscr EH}_{\xi} \pb{S} 2 n_\m u_\n \na^{[\m}\xi^{\n]}\eps_S
= -2 \bar n_\m \bar u_\n \na^{[\m}\xi^{\n]}\eps_S, \\
& i_\xi \ell^b = b\KB i_\xi \eps_{\cal T} = \f b2\KB \bar n^\m \xi^\n \eps_{\m\n\r\s}dx^\r\w dx^\s \pb{S}
% b\l \KB (\bar q \xi)\cdot n \, \eps_S= 
b \KB \xi\cdot \bar u\, \eps_S, \\
& I_\xi \vth^b \pb{S} - 2 n_\m u_\n \na^{(\m}\xi^{\n)} \eps_S = - 2 \bar n_\m \bar u_\n \na^{(\m}\xi^{\n)} \eps_S.
\end{align}
Adding up using the barred expressions, we find 
\begin{align}
H_\xi^b&=\int_S q^{\sscr EH}_{\xi}+i_\xi\ell^b-I_\xi\vth^b = 
\int_S -2\bar n_\m \bar u_\n ( \na^{[\m}\xi^{\n]} - \f b2 \KB \bar n^\m \xi^\n -\na^{(\m}\xi^{\n)} )  \eps_S 
\nn\\ &= -2\int_S \xi^\m\bar u^\n(\bar K_{\m\n} - \f b2\bar q_{\m\n}\bar K)   \eps_S.\label{tiltedBYb}
\end{align}
It is very similar to the orthogonal result, with the same dependence on the extrinsic curvature, but this time taken along $\bar u=\cosh\b n-\sinh\b u$: The normal to $\Si$ in the orthogonal case is replaced by its projection along the non-orthogonal boundary $\cal T$.

To manipulate further this expression, we use the decomposition \eqref{xiN} of the diffeomorphism, in terms of unbarred lapse and shift to facilitate the comparison with the canonical result below.
The restriction \eqref{xitang} to tangential diffeomorphisms implies a relation between the unbarred shift and lapse:
\be\label{tiltedshift}
\xi\cdot\bar n=0 \quad \Rightarrow \quad N^\m \bar n_\m = -N\sinh\beta.
\ee
The other scalar products give
\be\label{scalars}
\xi\cdot \bar u= -N\l,\qquad  \xi\cdot n= -N, \qquad \xi\cdot u =-N\l \sinh\b.
\ee
It is also convenient to decompose the shift vector as
\be\label{shiftdec}
N^\m = N\cdot u \, u^\m +\Nc^\m, \qquad N\cdot u=\xi\cdot u = -N\l\sinh\b,
\ee
and $\Nc^\m$ are the components tangent to the corner.
From the first of \eqref{scalars} it follows that the second integrand in \eqref{tiltedBYb} gives
\be\label{boosting1}
-\f b2\xi\cdot\bar u\bar K = \f b2 N\l \bar K.
\ee
For the first integrand, we use
\be\label{modric}
\bar u^\n\na_\m\bar n_\n = n^\n\na_\m u_\n+\p_\m\b,
\ee
which follows from the boost transformations \eqref{SO11},
to rewrite in terms of \eqref{xiN}
\begin{align}\label{boosting2}
\xi^\m\bar u^\n \bar K_{\m\n} &= \xi^\m \bar q^\r_\m \bar u^\n \na_\r \bar n_\n 
= \xi^\m (n^\n\na_\m u_\n+\p_\m\b) \nn\\ &= N(k-\na_\m u^\m) -N^\m u^\n K_{\m\n} +\dot\b,
\end{align}
where we used $\xi\cdot\bar n=0$ in the second equality, and \eqref{littlek} as well as $\xi=\p_t$  in the second line.
From $u^\m=\l (q \bar n)^\m$, see \eqref{uubar}, we have that
\be
\na_\m u^\m = \l \bar K+\l\na_{\bar u}\b+\l\sinh\b K.
\ee
Replacing this expression in \eqref{boosting2}, 
and using
\be\label{nabaru}
N\l\na_{\bar u} = \na_t - \Nc^\m\na_\m
\ee
which follows from \eqref{shiftdec}, we arrive at
\begin{align}\label{boosting3}
\xi^\m\bar u^\n \bar K_{\m\n} &=  N\Big(k-\l\bar K - \l\sinh\b (K- u^\m u^\n K_{\m\n} )\Big)-\Nc^\m (u^\n K_{\m\n} -\p_\m\b) \\\nn
& = N\Big(k - \l \KB -\l\sinh\b \,\jmath_\vdash \Big) + \Nc^\m (\jmath_{{\sscr S}\m} + \p_\m\b ),
\end{align}
where
\begin{align}
& \jmath_\vdash:= - u^\m u^\n \Pi_{\m\n} = K - u^\m u^\n K_{\m\n} = \g^{\m\n}K_{\m\n}, \\
& \jmath_{\sscr S}^\m := - \g^{\m\r} u^\n \Pi_{\r\n} = - \g^{\m\r} u^\n K_{\r\n},
\end{align}
are respectively the radial (or normal) and tangential momentum \cite{Brown:2000dz}. 
Adding up with \eqref{boosting1}, we find
\begin{align}\label{Htilted}
H_\xi^b
&= -2 \int_S \left(N\Big(k + \f{b-2}2 \l \KB -\l\sinh\b \,\jmath_\vdash \Big)
+ \Nc^\m (\jmath_{{\sscr S}\m} + \p_\m\b ) \right) \eps_S.
\end{align}

This formula provides the energy-momentum decomposition of the $b-$generalized Brown-York quasi-local charge with boundaries at non-orthogonal corners, in terms of \eqref{xiN} and using \eqref{tiltedshift}. We remark that it depends on the boundary Lagrangian via $b$, but not on $c$ and thus not on the corner Lagrangian. For $b=2$, it reproduces correctly $(4.6)$ of \cite{Brown:2000dz}.\footnote{To see the equivalence, we first observe that the notational translation from our paper to theirs is $(\b,\l,\tanh\b,u^\m,K_{\m\n}, k)\mapsto(-\th,1/\g,-v,n^\m,-K_{\m\n}, -k)$, 
and then recall their definitions $\bar N=N/\g, \bar\varepsilon=\g k - \g v \jmath_\vdash, \bar\jmath_{\sscr S}^\m=\jmath_{\sscr S}^\m-\g^{\m\n} \p_\n\th$.}
Notice that the contribution of the radial momentum becomes mixed with the energy because the radial shift component $N\cdot u$ is proportional to lapse, as a consequence of \eqref{tiltedshift}. To keep track of the momentum components separately from the energy, we can also rewrite \eqref{Htilted}as
\begin{align}\label{Htilted2}
H_\xi^b
&= -2 \int_S \left(N\Big(k + \f{b-2}2 \l \KB \Big) + N\cdot u \,\jmath_\vdash
+ \Nc^\m (\jmath_{{\sscr S}\m} + \p_\m\b ) \right) \eps_S.
\end{align}

%--------------------------------------------------------------
\section{Surface charges from canonical methods}\label{SecCan}
%--------------------------------------------------------------

In this section we review the canonical construction of \cite{Brown:1992br,Hawking:1996ww,Brown:2000dz}, and show that it extends to the mixed and Neumann boundary conditions. We consider directly the general case of non-orthogonal corners. In particular, we will show that $(i)$ the boundary terms recast the kinetic terms in the form  appropriate to the chosen boundary conditions, and $(ii)$ one reproduces the same expressions obtained with covariant phase space methods for orthogonal corners. 

We start from the $b$-generalized trace-K action \eqref{traceK}, and replace
\be
R\eps = L^{\sscr ADM}+2\na_\m(n^\m K-a^\m)\eps.
\ee
From Stokes theorem, 
\be
\int_M \na_\m(n^\m K-a^\m)\eps = \int^{\Si_1}_{\Si_0}K\eps_\Si + \int_{\cal T} (\sinh\b \, K - \bar n\cdot a) \eps_{\cal T}.
\ee
Therefore, \eqref{traceK} can be rewritten as follows,
\begin{align}\label{S2}
S=\int L^{\sscr ADM} + (2-b)\int^{\Si_1}_{\Si_0} K\eps_\Si+\int_{\cal T} ( b \KB +2 \sinh\b \, K- 2 \bar n\cdot a) \eps_{\cal T} + c \int^{S_1}_{S_0}\b \eps_S.
\end{align}

The first two integrals above give the bulk terms on the space-like slices.
The ADM Lagrangian density is
\begin{align}
& \tl L^{\sscr ADM} := N\sqrt{q}({\cal R}+K_{\m\n}^2-K^2) 
= \tl\Pi^{\m\n}\dot q_{\m\n}-N \tl\cH - N^a\tl\cH_\m -2\p_\m(\tl\Pi^{\m\n} N_\n),
\end{align}
where
\be 
\tl\cH := \f1{\sqrt{q}}(\tl\Pi_{\m\n}^2-\f12\tl\Pi^2)-\sqrt{q} {\cal R},
\qquad  \tl\cH_\m:=-2\tl q_{\m\n}D_\r\Pi^{\n\r}
\ee
are the Hamiltonian and spatial diffeo constraints, and the boundary term is the one giving rise to the ADM momentum 
\be\label{Padm}
P(\vec N):=2\int_S \Pi^{\m\n}u_\m N_\n \,\eps_S.
\ee
The second integral can be rewritten as the spacetime integral of $(2-b)\p_t(K\sqrt{q})= \tfrac12(b-2) \p_t\tl\Pi$. This combines with the kinetic term of the ADM Lagrangian, giving
\be
\tilde{\Pi}^{\m\n}\dot q_{\m\n} + \f{b-2}2 \dot{\tl\Pi} = \left\{\begin{array}{ll}
\tilde{\Pi}^{\m\n}\dot q_{\m\n} & b=2 \vspace{.3cm} \\ - q_{\m\n}\dot{\tilde{\Pi}}^{\m\n} & b=0 \vspace{.35cm} \\ \tl P^{\m\n}\dot q_{\m\n} +\f43\sqrt{q}\dot K \qquad & b=2/3
\end{array}\right.
\ee
The resulting polarization is
in agreement with the discussion in Section~\ref{sec:mixed}.
In other words,  different choices of $b$ correspond to   different boundary conditions, and the role of the second integral in \eqref{S2} is to adjust the kinetic term to the chosen coordinate-momentum pair.  

As we will see shortly, the Hayward term plays the same role for the boundary kinetic term.

The last two integrals in \eqref{S2} will reveal the value of the sourface term corresponding to the energy. To make this explicit, we use first Stokes theorem to rewrite
\begin{align}\label{HHStokes1}
\int^{S'}_S \b\,\eps_S =\int_{\cal T} \bar D_{\m}(\b \bar u^\m)\,\eps_{\cal T} = \int dt\int_S d^2x \, \p_t(\b\sqrt{\g}),
\end{align}
where in the last step we decomposed ${\cal T}$ into its $t$ foliation with space-like leaves $S$.
Then, using the relation \eqref{littlek} between extrinsic curvatures, the last two integrals in \eqref{S2} give
\begin{align}\label{ellT}
&\int_{\cal T}\ell_{\cal T}:= \int_{\cal T} \left(2 \cosh\b \, k + (b-2) \bar K + (c-2)\na_{\bar u}\b + c \b \bar D_\m \bar u^\m \right) \eps_{\cal T}. 
\end{align}
However, this surface term is not yet the contribution to the energy, because it is not in Hamiltonian form: both $\na_{\bar u}$ and the Hayward term contain time derivatives, and a Legendre transform is needed to read the correct Hamiltonian. This procedure was explained in \cite{Brown:2000dz} for $c=2$, and in \cite{Harlow:2019yfa} for $c=0$.\footnote{This point was overlooked in the first preprint version of this paper, and led us to the erroneous conclusion that there was a discrepancy between the covariant and canonical charges for non-orthogonal corners.} 
To make time derivatives explicit, we use \eqref{nabaru} and the second equality in \eqref{HHStokes1}. Then \eqref{ellT} gives
\begin{align}\label{ellT2}
&\int_{\cal T}\ell_{\cal T} = \int dt\int_{S} d^2x \left[(c-2)\sqrt\gamma \dot\beta + c\beta\skew{16}\dot{\sqrt\gamma} + \sqrt{\gamma} \Big( N(2k + (b-2)\l {\bar K}) +2 \Nc^\m\p_\m\b \Big)\right],
\end{align}
where we used the relation \eqref{dets} between determinants to replace $\eps_{\cal T}$ with $N\l\sqrt{\g}$.
The first two terms give a polarization of the phase space according to the chosen boundary conditions: 
 \be
(c-2)\sqrt\gamma \dot\beta + c\beta\skew{16}\dot{\sqrt\gamma} =\begin{cases}
-2\sqrt\gamma\dot\beta \qquad & c=0\\
2\beta\skew{16}\dot{\sqrt\gamma} \qquad &c=2
\end{cases}
\ee

For $c=2$, $\ell_{\cal T}$ contains the boundary kinetic term $2\b\p_t{\sqrt\g}$, which is in Dirichlet form $pdq$. The boundary momentum is then
$p_\g=2\b$, and the Legendre transform gives
\begin{align}
& \ell_{\cal T} = p_\g \skew{16}\dot{\sqrt\gamma} -{\cal H}_{\cal T}, \\
\label{HT} & {\cal H}_{\cal T} = -2\sqrt{\g}\Big[N\Big(k+\f{b-2}2\l\bar K\Big) + \Nc^\m\p_\m\b\Big].
\end{align}
For $c=0$, the boundary kinetic term is instead $-2{\sqrt\g}\p_t\b$, which is of Neumann/York form $-qdp$, according to the Table at the bottom of Section 2. The momentum is now $p_\b=-2\sqrt\g$, and the Legendre transform gives
\begin{align}
& \ell_{\cal T} = p_\b\dot\b-{\cal H}_{\cal T}, 
\end{align}
with precisely the same Hamiltonian \eqref{HT} again. This is the correct surface contribution to the Hamiltonian, and it is  independent of $c$, as in  the covariant phase space result. The only effect of the Hayward term is to adjust the boundary kinetic term from Dirichlet to Neumann/mixed form. The total action is thus
\begin{align}\label{S3}
S&=\int dt \left[\int_\Si d^3x \left(\tilde{\Pi}^{\m\n}\dot q_{\m\n} + \f{b-2}2 \dot{\tl\Pi} -N \tl\cH - N^a\tl\cH_\m \right) \right.\\\nn
& \left.\qquad\qquad +\int_Sd^2x\Big((c-2)\sqrt\gamma \dot\beta + c\beta\skew{12}\dot{\sqrt\gamma} - \sqrt{\g} \Big( E(N,\b) +P(\vec N,\b)\Big)\right],
\end{align}
where
\begin{align}\label{canE}
& E(N,\b)= -2 \int_{S} N\left(k + \f{b-2}2\l {\bar K} \right) \eps_S, \\
& P(\vec N,\b) = 2 \int_{S} \left( \Pi^{\m\n}u_\m N_\n -\Nc^\m\p_\m\b\right) \eps_S
= - 2 \int_{S} \Big( N\cdot u \,\jmath_\vdash + \Nc^\m(\jmath_{{\sscr S}\m} + \p_\m\b)\Big) \eps_S.
\end{align}
The charges coincide perfectly with \eqref{Htilted2}, for all values of $b$ and $\b$.
The canonical and covariant results thus match for all boundary conditions considered.

As a final remark, let us say that a discrepancy for non-orthogonal corners could have been expected, since a corner Lagrangian is needed for the variational principle and it was not included in the application of \eqref{HamC}. The reason why it does not happen is that the corner term can always be considered as part of the space-like boundary instead of the time-like boundary, as thus it enters the specification of the state, and not of the phase space \cite{Harlow:2019yfa}. We confirm this, and what we have seen is that by keeping track of the space-like boundary terms, one can read the form of the boundary kinetic terms associated with the  chosen boundary conditions.

%--------------------------------------------------------------
\section{Conclusions}
%--------------------------------------------------------------

We have applied the procedure outlined in \cite{Wald:1999wa,Harlow:2019yfa,Freidel:2020xyx}, and more precisely the formula given in \cite{Freidel:2020xyx},  to obtain Hamiltonians from conservative boundary conditions from covariant phase space methods, extending the Dirichlet analysis of \cite{Iyer:1995kg,Harlow:2019yfa,Freidel:2020xyx} to the case of mixed and Neumann boundary conditions. We have provided explicit formulas for the charges, showing how the charges and in particular the energy depend on the choice of boundary conditions. We have then compared these formulas with the analogue calculations that can be done using canonical methods. We found a perfect matching for 
both 
orthogonal and non-orthogonal corners, confirming the dependence of the energy on the boundary conditions, and discussed the role of the Hayward corner Lagrangian in settling the boundary kinetic term to the form associated with the boundary conditions chosen.
We would like to highlight three implications of our results. First, they provide additional support for the prescription of \cite{Freidel:2020xyx}, by showing that it reproduces the canonical results for different boundary Lagrangian, and how to amend it in the extension to non-orthogonal corners. 
Second, they bring more attention to the charges associated with York's mixed boundary conditions \cite{An:2021fcq}, which have been argued to give a better posed initial-boundary value problem \cite{Anderson:2006lqb,Anderson:2010ph}. Third, they will hopefully encourage discussions about the dependence of the energy on the boundary conditions.

This dependence was anticipated in \cite{Iyer:1995kg}, and we have provided a quantitative analysis thereof, given by \eqref{charge} or \eqref{chargelapse} and \eqref{Htilted}. The values obtained for the energy with different boundary conditions are summarized in Table~\ref{Table2} below. For shortness of notation, we restrict to orthogonal corners. The quasi-local value of the energy is given prior to renormalization, whereas the value given for the Kerr spacetime is the asymptotic energy at spatial infinity, after renormalization, as in \eqref{Hren}.
\begin{table}[h] \begin{center}
  \begin{tabular}{l|c|c|c|c}  
\emph{boundary conditions} & \emph{quantity held fixed} & \emph{value of $b$} & \emph{quasi-local energy} 
& \emph{Kerr (renormalized) } \\ \hline
Dirichlet & $q_{\m\n}$ & 2 & $k$ & $M$\\
York & $(\hat q_{\m\n}, K)$ & $2/3$ & $k-2\KB/3$ &$2 M/3$\\
Neumann & $\tl\Pi^{\m\n}$ & 0 & $k-\KB$ &$M/2$
\end{tabular}
  \caption{\label{Table2} \small{\emph{Different values of the energy computed as the generator of time-diffeomorphisms at conservative boundaries.
  A feature that inevitably catches the eye is that choosing boundary conditions with fewer components of the induced metric fixed, and more components of the momentum fixed, produces a smaller value of the energy for time-like boundaries with positive extrinsic curvature, as in the Kerr example.
One may speculate an interpretation for this by saying that holding the momenta fixed at the boundary instead of the configuration variables, energy is being stored there and removed from the system. }}  \label{TabEnergy}}
 \end{center}   
 \end{table}

It is important to stress that this dependence is a special feature of the role of boundary terms in field theory and general relativity in particular: in a finite-dimensional system, the energy  does \emph{not} depend on the choice of boundary conditions.\footnote{Think for example of the Dirichlet and Neumann Lagrangians for a point particle, 
$L^{\sscr D}=\dot x^2/2$ and $L^{\sscr N}=-x\ddot x/2 = L^{\sscr D} +d\ell$, with $\ell=-x\dot x/2.$ To compute the energy in the latter case, one can use the method of Ostrogradsky (see e.g. \cite{Woodard:2015zca} for a modern description) and define two momenta 
\be\nn
p_1:=\f{\p L^{\sscr N}}{\p \dot x}-\f{d}{dt}\f{\p L^{\sscr N}}{\p \ddot x} =\f12\dot x, 
\qquad p_1:=\f{\p L^{\sscr N}}{\p \ddot x} =-\f12x.
\ee
The energy is then given by
\be\nn
E^{\sscr N}:=p_1\dot x+p_2\ddot x-L^{\sscr N} = \f12\dot x^2\equiv E^{\sscr D},
\ee
matching the standard expression obtained with $L^{\sscr D}$.}
Whether the construction of gravitational surface charges should be amended to achieve the same independence, so that for instance one always finds the Newtonian mass $M$ in the case of Kerr, or whether there is a deeper physical meaning in a notion of canonical energy whose value depends on the boundary conditions chosen, is something that we believe needs further discussions in future work.
If consolidated, this feature would stress the non-trivial role that the boundary representation plays in general relativity, something already observed when changing variables and formulations (see e.g. discussions in \cite{Freidel:2015gpa,DePaoli:2018erh,Freidel:2020xyx,Freidel:2020ayo,Godazgar:2020kqd,Oliveri:2020xls,Geiller:2021gdk}), and here found when changing boundary conditions. It can only be expected that such a dependence at the classical level would be even more relevant in the quantum theory. 
%

%--------------------------------------------------------------------------------------------------
\subsection*{Acknowledgments}
%--------------------------------------------------------------------------------------------------
We thank David Hilditch for discussions on the initial boundary value problem, and Laurent Freidel, Marc Geiller, Roberto Oliveri,  Daniele Pranzetti, Carlo Rovelli and Wolfgang Wieland for discussions on surface charges.  We are grateful to an anonymous referee for explaining to us how the boundary Legendre transform is necessary in order to obtain the correct match between covariant and canonical results with non-orthogonal corners. We thank Michael Anderson and Zhongshan An for useful feedback on our draft.

%---------------------------------------------------
\appendix
%---------------------------------------------------

%----------------------------------------------------------------------------
\section{Notations and kinematics}\label{AppA}
%----------------------------------------------------------------------------

With reference to the notation spelled out in Fig.~\ref{Fig1} and Table~\ref{tab:defn}, we have
\begin{align}
& \Si: \qquad 
n^2=-1, \qquad q_{\m\n} = g_{\m\n}+n_\m n_\n, \qquad u\cdot n=0, \qquad u^2=1 \\
& {\cal T}:\qquad 
\nB^2=1, \qquad \qB_{\m\n} = g_{\m\n}- \nB_\m \nB_\n, \qquad \uB\cdot\nB=0, \qquad \uB^2=-1 \\
& S: \qquad \g_{\m\n}=q_{\m\n}-u_\m u_\n = g_{\m\n}+n_\m n_\n -u_\m u_\n 
= g_{\m\n} - \bar n_\m \bar n_\n +\bar u_\m \bar u_\n.
\end{align}
Boost relations between normals:
\begin{align}
& n\cdot \nB = \sinh\b, \qquad u\cdot \nB = \cosh\b, \qquad \uB\cdot n = - \cosh\b, \qquad \l:=(\cosh\b)^{-1}\\\label{SO11}
& \vet{n}{u} = \mat{\cosh\b}{\sinh\b}{\sinh\b}{\cosh\b}\vet{\uB}{\nB}, \qquad \vet{\uB}{\nB}= \mat{\cosh\b}{-\sinh\b}{-\sinh\b}{\cosh\b} \vet{n}{u} \\
& u^\m = \l q^\m_\n \nB^\n =  \l(\nB^\m+\sinh\b n^\m), \qquad 
 \uB^\m = \l  \qB^\m_\n n^\n = \l  (n^\m-\sinh\b \nB^\m) \label{uubar}
 \end{align}
Induced derivatives:
\be
D_\m (q_{\n\r}f^\r)=q^\s_\m\na_\s(q_{\n\r}f^\r), \qquad \bar D_\m (\bar q_{\n\r}f^\r)=\bar q^\s_\m\na_\s(\bar q_{\n\r}f^\r)
\ee
Extrinsic curvatures:\footnote{For the reader familiar with the Brown-York papers, it is useful to recall that they use opposite signs in the definition of the extrinsic curvatures.}
\begin{align}
& K_{\m\n}:= q^\r_\m \na_\m n_\n, \qquad K=\na_\m n^\m, \qquad a^\m=n^\n\na_\n n^\m \\
& \KB_{\m\n}:= \qB^\r_\m \na_\m \nB_\n, \qquad \KB=\na_\m \nB^\m, \qquad \bar a^\m=\nB^\n\na_\n \nB^\m \\
& k_{\m\n}:= \g_\m^\r D_\r u_\n = \g_\m^\r q_\n^\s D_\r u_\s, 
\qquad \bar k_{\m\n}:=\g_\m^\r \bar D_\r \bar u_\n  \\
& k=g^{\m\n}k_{\m\n} = \g^{\m\n}\na_\m u_\n = \na_\m u^\m +n^\m n^\n \na_\m u_\n = 
\l (\KB + \sinh\b K+ \na_{\bar u}\beta - \nB\cdot a) \label{littlek} \\
& \bar k=g^{\m\n}\bar k_{\m\n} = \g^{\m\n}\na_\m \bar u_\n = \na_\m \bar u^\m - \bar n^\m \bar n^\n \na_\m \bar u_\n = 
\l  (K - \sinh\b \bar K+ \na_{u}\beta + n\cdot\bar a)\label{littlekbar}
\end{align}

Foliations: For the section on the canonical analysis, we take the boundaries $\Si$ and $\cal T$ to be part of foliations defined by the level sets of two scalar fields $\vphi^0(x)$ and $\vphi^1(x)$ respectively. Without loss of generality we can adapt our coordinates such that $\vphi^0=t$ and $\vphi^1=r$ in spherical coordinates, so that the corners defined by the intersections of the two foliations are spheres parametrized by $\th$ and $\phi$. In these adapted coordinates, the presence of a non-orthogonal corner is directly parametrized by one of the metric components, 
\be\label{betag}
n=-\f1{\sqrt{-g^{tt}}}dt, \qquad \nB=\f1{\sqrt{g^{rr}}}dr,\qquad \sinh\b=-\f{g^{tr}}{\sqrt{-g^{tt}}\sqrt{g^{rr}}}.
\ee
This identification can be used to provide a bulk extension of the function $\b$.
We further have
\be\label{dets}
 \sqrt{-g} = N\sqrt{q}, \qquad \sqrt{-\bar q} = N\l \sqrt{\g},
\ee
in terms of the ADM variables $N=-1/{\sqrt{-g^{tt}}}$, and 
\be
q_{ab}=g_{ab}=\mat{q_{rr}}{q_{rA}}{}{\g_{AB}},
\qquad
\bar q_{\bar a\bar b}=\mat{-N^2+q_{ab} N^aN^b}{q_{Ab}N^b}{}{\g_{AB}}.
\ee
In this set-up, the orthogonal corner case corresponds to a partial gauge-fixing in which one component of the shift vector vanishes, $N^r=0$. 
It is also possible to consider a more general set-up, in which the time-like boundary is not a level set of one of the coordinates. In this case one can describe both orthogonal and non-orthogonal cases without gauge fixing.

\subsection{Volume forms and pull-backs} \label{AppConv}

We denote the volume 4-form by
\be
\eps:=\f1{4!}\eps_{\m\n\r\s}~dx^\m\w dx^\n\w dx^\r\w dx^\s, \qquad \eps_{0123}:=\sqrt{-g}.
\ee

The induced volume 3-form on $\Si$ with normal $n_\m$ such that $n^2=s:=\pm1$, is
\be
\eps_\Si\equiv d\Si := i_n \eps =s n^{\m}d\Si_{\m}, \qquad \eps=s n\w \eps_\Si, \qquad i_n\eps_\Si
\ee
where $d\Si_\m:=sn_\m d\Si =\sqrt{-sq}~d^3y$ is the  oriented volume element in the conventions  of \cite{Lehner:2016vdi}.
Accordingly, the pull-back of a 3-form on $\Si$ reads
\begin{align}\label{3dpb}
\th \eqSi  \f s{3!}\th_{\m\n\r} \eps^{\m\n\r\s} n_\s d\Si = -\f s{3!}\th_{\m\n\r} \eps_\Si^{\m\n\r}d\Si
= \f1{3!}\th_{\m\n\r} \eps^{\m\n\r\s} d\Si_\s =: \th^\m d\Si_\m.
\end{align}

The induced volume 2-form on $\p\Si$ with normal $u_\m$ such that $u\cdot n=0$ and $u^2=-s$, is
\be
\eps_S\equiv dS:=-si_{u}d\Si = -n^{\rho}u^{\s}dS_{\r \s},  \qquad \eps_\Si=u\w \eps_S, \qquad \eps=s n\w u\w \eps_S,
\ee
where $dS_{\r \s} =2n_{[\rho}u_{\s]}dS$ is the oriented surface element with both outgoing normals. 
Accordingly, the pull-back of a 2-form $\a$ on $\p\Si$   is
\begin{align}
\a &\stackrel{\p\Si}= -\f s{2}\a_{\m\n} \eps^{\m\n\r\s} n_\r u_\s dS =-\f s2 (\star\a)^{\m\n} dS_{\m\n}.
\end{align}

To give some explicit intuition about these conventions, for $\Si$ and $\cal T$ defined in Minkowski space respectively by $t=const.$ and $r=const.$, we would have
\[
\eps_\Si = dr\w d\th\w d\phi, \qquad \eps_{\cal T}= -dt\w d\th\w d\phi,
\]
with pull-backs on their boundaries
\[
i_{\p_r}\eps_\Si = d\th\w d\phi =\eps_S, \qquad i_{\pm\p_t}\eps_{\cal T}= -dt\w d\th\w d\phi = \mp d\th\w d\phi =\mp\eps_S.
\]

%----------------------------------------------------------------------------
\bibliographystyle{JHEPs}
\bibliography{biblio}
%----------------------------------------------------------------------------

%--------------------------------------------------------------
\end{document}